# High Temperature Multiferroic State in $RBaCuFeO_5$ (R= Y, Lu and Tm)

Yuji Kawamura, Takahiro Kai, Erika Satomi, Yukio Yasui, Yoshiaki Kobayashi, and Masatoshi Sato *

*Department of Physics, Division of Material Science, Nagoya University, Furo-cho, Chikusa-ku Nagoya 464-8602 Japan*

Magnetic and/or dielectric behaviors have been studied for $YBaCuFeO_5$, $LuBaCuFeO_5$ and $TmBaCuFeO_5$, which are known as members of oxygen-defect ordered perovskite systems $RBaCuFeO_5$ (R=lanthanides Ln and other trivalent elements) and have two magnetic transitions: The magnetic structure of the high temperature($T$) ordered phase is basically antiferromagnetic, and with decreasing $T$, a modulated magnetic component superposed on the antiferromagnetic moments appears at the second transition, where the ferroelectricity also appears (multiferroic). Here, we have found that all the systems studied here exhibit multiferroic behavior in the low $T$ phase, and that $TmBaCuFeO_5$ becomes multiferroic at a temperature as high as the melting point of ice, which is, to our knowledge, the highest value ever found in zero magnetic field.

Multiferroics that exhibit ferroelectric and magnetic simultaneous transitions have been actively studied with interest in both fields of basic and applied sciences.[1) ] $LiVCuO_4$[2, 3)] and $LiCu_2O_2$[4-7)] with quasi one-dimensional $CuO_2$ chains ($CuO_2$ ribbon chains) formed of the edge-sharing $CuO_4$ square plaquettes are known as example systems in the former field: They are expected to exhibit new quantum phenomena induced by the coupling between magnetism and electric polarization. Their Cu spins ($s$ =1/2) exhibit helical ordering with the modulation vector $\boldsymbol{Q}$ along the chains, and a spontaneous electric polarization $\boldsymbol{P}$ ($\propto \boldsymbol{Q}\times\boldsymbol{e}_3$) appears as derived by theoretical consideration reported in refs. 8 and 9 ( The unit vector $\boldsymbol{e}_3$ is defined parallel to the helical axis). In addition to this system, we have also found that $Na_2Cu_3(GeO_3)_4$, which has $Cu_3O_8$ clusters formed of three edge-sharing $CuO_4$ square planes, is a multiferroic.[10)]

Systems with large magnetic and ferroelectric polarizations attract much interest from the view point of technical application, because their (electric polarizations)/(magnetic ordering patterns) can be controlled by the external magnetic/electric fields.[1)] For example, perovskite Mn oxides such as $TbMnO_3$ and $DyMnO_3$ have been studied very extensively, and it has been experimentally proved that the electric polarization $\boldsymbol{P}$ can be controlled by the external magnetic filed $\boldsymbol{H}$.[11, 12)]

Then, from the view points of applications, multiferroic systems above room temperature are quite desirable. However, there is no example system exhibiting multiferroic behavior above room temperature (RT), although Zn ferrite system is known to be multiferroic above RT under applied magnetic field.[13)] Here, stimulated by a report by Kundys *et al.*[14)] that $YBaCuFeO_5$ is a multiferroic with a rather high transition temperature of ~230 K, we have adopted a series of systems $RBaCuFeO_5$ (R=lanthanides Ln and other trivalent elements) to search for high temperature multiferroics, and in the present paper, we report results of magnetic and dielectric or ferroelectric measurements carried out for R=Y, Lu and Tm. We have found that for R=Tm, the ferroelectric polarization $P$ appears with decreasing $T$ at a temperature as high as the melting point of ice, which is, to our knowledge, the highest transition temperature to a multiferroic phase in zero magnetic field.

Polycrystalline samples of $RBaCuFeO_5$ (R=Y, Lu and Tm) were prepared by a solid reaction: Mixtures of $R_2O_3$, $BaO_2$ (or $BaCO_3$), CuO and $Fe_2O_3$ powders with the molar ratios 1:2:2:1 were ground and pelletized, and the pellets were heated in air to 1000 ºC, kept at the temperature for 24 h, and cooled down to room temperature. Then, the processes of regrinding and heat treatments were repeated two or three times, and finally, the pellets were annealed in Ar gas flow at 500 °C for 10 h. We confirmed that the samples were single phase by X-ray measurements with CuKα radiation. In Fig. 1, the X-ray diffraction pattern obtained for R=Tm is shown, for example, together with the one calculated for R= Y.[15)] We have confirmed that all the systems studied here have the structures schematically shown in the right part of Fig. 1. They are tetragonal (space group P4mm**)**. Cu and Fe atoms are considered to be ordered as shown in the figure.[15)] (The existence of this ordering has been confirmed by the magnetic structure analysis in the antiferromagnetically ordered phase for R=Y,[16)] though it is not a subject of the present paper.)

The magnetic susceptibility was measured using a SQUID magnetometer in the temperature $T$ range from 5 to 650 K. The $T$ dependence of the dielectric susceptibility ε was studied by measuring the capacitance $C$ of circular plate samples (typical sizes of 9 mmϕ× ~2.5 mm thick) with Au evaporated electrode films using a capacitance bridge (Andeen Hagerling) with the frequency of 1 kHz. The $T$ dependence of the electric polarizations $P$ was obtained from the data of the pyrocurrent $I_p$. The measurements of $I_p$ were carried out by the following procedures: First, samples were cooled under applied voltage $V$ between the electrodes, and after $V$ was set at 0, $I_p$ was measured with varying $T$ from the lowest temperature studied here (~77 K) using an electrometer (Keithley 6517A), where the time $t$ was also measured. The d$T$/d$t$ was, roughly ~20 K/min. (We had to wait ~20 minutes at the lowest temperature to minimize the spurious current.)

Figures 2(a)-2(d) show the $T$ dependences of the magnetic susceptibility χ, dielectric susceptibility ε, pyroelectric current $I_p$ and electric polarization $P$, respectively, obtained for a polycrystalline sample of

*corresponding author (e43247a@nucc.cc.nagoya-u.ac.jp)

$YBaCuFeO_5$. In Fig. 2(a), we can find peaks of $\chi$ at temperatures of $T_{N1}$~455 K and $T_{N2}$~180 K at $H$=1 T ($T_{N2}$~189 K at $H\to 0$). If we simply expect the relation $I_p \propto (dP/dT)\cdot(dT/dt)$, $I_p$ can roughly be considered to be proportional to $dP/dT$, because $(dT/dt)$ does not change very sensitively with $T$ in the $T$ region of nonzero $I_p$. To the data curves in Figs. 2(c) and 2(d), the electric voltages applied between the electrodes during the sample cooling to remove the ferroelectric domains are attached. Although these temperatures do not accurately coincide with the values reported in ref. 14, characteristic features of $\chi$ are reproduced, and we can say that with decreasing $T$, a magnetic transition to the antiferromagnetic state[15, 16] at around $T_{N1}$ and the second magnetic transition to an modulated structure at around $T_{N2}$ take place as was found in refs. 15 and 16. Although this modulated structure has not been determined yet, we know that neutron superlattice reflections appear at the reciprocal points of ($h$/2, $k$/2, $l$/2±δ) with odd values of $h$ $k$ $l$ and δ~0.1. (Strictly speaking, the ordering temperature $T_{N2}$' should be determined as the peak temperature of $d\chi/dT$, which is lower than $T_{N2}$ by ~10 K in the limit of zero magnetic field, as obtained from the data shown in Fig. 3(b) for $LuBaCuFeO_5$, for example. We indicate the $T_{N2}$' values by the broken lines in Figs. 2(b)-2(d), and hereafter we use $T_{N2}$' as the transition temperature.) In Fig. 2(b), we find that an anomaly of ε at ~$T_{N2}$', and that the electric polarization $P$ appears with decreasing $T$ at ~ $T_{N2}$' (see Fig. 2(d)). These results indicate that the magnetic moments couple with the electric polarization $P$,[15] and the existence of the modulated magnetic structure below $T_{N2}$' is consistent with the relation $\boldsymbol{P}\propto\boldsymbol{Q}\times\boldsymbol{e}_3$.[8, 9] Therefore, we first try to understand the results on the basis of this model without having precise knowledge on the modulated magnetic structure.

Let us consider that in the low $T$ ordered phase, sinusoidally modulated components of the magnetic system, $M_X$ and $M_Y$, appear along two orthogonal directions, say $X$ and $Y$ directions, with the modulation amplitudes $M_X^0$ and $M_Y^0$, respectively, forming a helical component with the amplitude $M^0$. (Note that the $X$, $Y$ directions are not necessarily coincide with the crystal axes, and that if $M_X^0 = M_Y^0$, we have the relations $M^0 = M_X^0 = M_Y^0$), Then, if we expect the mean field behavior, $M_X^0$ and $M_Y^0$ are proportional to $(T_{N2}'-T)^{0.5}$, the polarization $P$ $(\propto M_X^0 \cdot M_Y^0)$ [8, 9] $\propto(T_{N2}'-T)$ holds in the $T$ region of $(T_{N2}'-T)/T_{N2} \ll 1$, and $I_p$ which is roughly proportional to $|dP/dT|$ should be largest at ~$T_{N2}$' with rather flat $T$ dependence near the transition temperature. However, $I_p$ shown Fig. 2(c) does not exhibit such the behavior. Instead, it has a maximum value at temperature $T_0$, much lower than $T_{N2}$' and decreases as $T$ approaches $T_{N2}$'.

In Figs. 3(a) and 3(b), $T$ dependences of the electric polarization $P$ and temperature derivative of the magnetic susceptibility, $d\chi/dT$, are shown. They are obtained for a polycrystal sample of $LuBaCuFeO_5$ and an aligned sampler of $LuBaCuFeO_5$ (in stycast), respectively. In the inset of Fig. 3(b), the $T$ dependence of $\chi$ is also shown. We find that $T_{N2}$' ~178 K (at $H\to 0$) and $T_{N2}$ ~ 184 K (at $H$=1 T). The appearance of the superlattice reflections with incommensurate wave vector at round $T_{N2}$',[17] with decreasing $T$, indicates that the modulated structure exists in the low $T$ phase. We could observe no anomaly in the ε-$T$ curve, possibly because the absolute value of ε was large. For this system, we realize again that $P$ appears at around $T$~$T_{N2}$' with decreasing $T$, and the critical exponent α used to describe the relation $P\propto(T_{N2}'-T)^\alpha$ in the $T$ region near $T_{N2}$' seems to be larger than unity. As in the case of $YBaCuFeO_5$, the temperature of $I_p$ maximum, $T_0$ of $LuBaCuFeO_5$ is significantly lower than $T_{N2}$'. Between $T_0$ (~120 K) and $T_{N2}$', $|I_p|$ or $|dP/dT|$ exhibits a significant decrease, with increasing $T$, in contrast to the $T$ dependence expected for the mean field behavior stated above.

Figure 4 shows the $T$ dependence of the electric polarization $P$ obtained for $TmBaCuFeO_5$ by measuring the pyroelectric current $I_p$. For this system, we could not see such the peaks in the $\chi$-$T$ curve, as those found in YBaCuFeO5 and LuBaCuFeO5 at $T_{N1}$ and $T_{N2}$, because $Tm^{3+}$ ions have large magnetic moments. We could not observe the anomaly in the ε-$T$ curve, either, for the same reason as in the case of $LuBaCuFeO_5$, although the existence of the low temperature incommensurate phase has been reported.[17] The $T$ dependence of $P$ is very similar to that observed for $YBaCuFeO_5$. An interesting point is that the finite polarization $P$ appears, with decreasing $T$ at a temperature almost equal to the melting point of ice, which is, to our knowledge, the highest critical temperature of multiferroic transitions in zero magnetic field.

For well known multiferroic systems, $TbMnO_3$[12] and $MnWO_4$,[18] for example, the $T$ dependence of $P$ seems to be roughly consistent with the relation $P\propto (T_{N2}'-T)^{0.5}$. In their cases, a sinusoidal modulation of ordered magnetic moments with $M_X^0 \neq 0$ and $M_Y^0=0$ appears first with decreasing $T$, and at lower temperature $T_{N2}$', where $M_X^0$ approaches a $T$-insensitive value, $M_Y^0$ with usual $T$ dependence appears, inducing nonzero $P$ $(\propto M_X^0\cdot M_Y^0) \propto M_Y^0 \propto (T_{N2}'-T)^{0.5}$. In contrast, to understand the $T$ dependence of $P$ of the present systems, the relation $P$ (or $M_X^0\cdot M_Y^0)\propto(T_{N2}'-T)^\alpha$ with $\alpha \geq 1$ have to be used. To explain this relation by the same theoretical model as that used above, we consider a simultaneous appearance of $M_X^0$ and $M_Y^0$, each of which is proportional to $(T_{N2}'-T)^{\alpha/2}$ $(\alpha\geq 1)$ and superposed on the antiferromagnetically ordered moments. This type of $T$ dependence of $M_X^0$ and $M_Y^0$ can be found, for example, in FeGe with hexagonal structure.[19] The system exhibits a transition, with decreasing $T$, to a magnetic structure with incommensurate helical components, where both modulation amplitudes, $M_X^0$ and $M_Y^0$ superposed on an antiferromagnetic structure are proportional to $(T_{N2}'-T)^{\alpha/2}$ $(\alpha>1)$. We also know a similar superposition of helical components with α significantly larger than 1 on the ferromagnetic moment system of $Cu_{1-x}Zn_xCr_2Se_4$.[20] If the present systems exhibit similar characteristics of magnetic behavior to those found in the above systems, the $T$ dependence of $P$ observed here can be understood. However, the $T$ dependence of the neutron scattering intensities from the modulated magnetic structure reported previously for $YBaCuFeO_5$[15, 21] does not seem to have such characteristics found for FeGe and $Cu_{1-x}Zn_xCr_2Se_4$. Therefore, we think that the alternative explanation is more persuasive: As $T$ is raised, ferroelectric domains are formed, because no external fields are applied during the measurement of $I_p$. It can consistently explain the appearance of the maximum of $I_p$ at $T_0$ significantly lower than $T_{N2}$'. Note that though the reconstruction of the domain structure induces the change of $I_p$, but does not induce the anomaly in the ε-*T* curve.

Two other questions also arise here: One is why the transition temperatures of the present system are so high. Because the modulation vector of the magnetic ordering is along the *c* axis, it is not plausible that the large exchange interaction among $Cu^{2+}$ moments expected within a *c*-plane is important. We presume, therefore, that the large magnetic moments of $Fe^{3+}$ and isotropic nature of both the $Fe^{3+}$ and $Cu^{2+}$ spins are important for realizing the non-collinear modulated structure (possibly helical structure) of the systems. If it is correct, for the search for new multiferroic systems with higher transition temperatures, it is important to substitute $Cu^{2+}$ with $Mn^{2+}$ ions, for example, which have large and isotropic moments. Substitutions of $Ba^{2+}$ with different size ions such as $Sr^{2+}$ or $Ca^{2+}$ would also be hopeful, because a slight change of the atomic distances may induce the enhancement of the exchange interactions among the magnetic moments. The other question is why the absolute values of *P* observed for the present systems at low *T* are smaller than the value reported by Kundys *et al.*[14] for $YBaCuFeO_5$. On this point, we do not find firm reasons at this moment.

In summary, we have studied magnetic and/or dielectric behaviors of $RBaCuFeO_5$ (R=Y, Lu and Tm), and found that all these systems exhibit multiferroic properties. In particular, the transition temperature of $TmBaCuFeO_5$ have been found to be as high as the melting point of ice, which is, to our knowledge, the highest ever found in zero magnetic field. It encourages us to search for a room temperature multiferroic system among similar systems.

## Acknowledgements

This work is supported by Grants-in-Aid for Scientific Research from the Japan Society for the Promotion of Science (JSPS) and Grants-in Aid on priority areas from the Ministry of Education, Culture, Sports, Science and Technology.

1) S.-W. Cheong and M. Mostovoy: Nature Mater. **6** (2007) 13.
2) Y. Naito, K. Sato, Y. Yasui, Y. Kobayashi, Y. Kobayashi, and M. Sato: J. Phys. Soc. Jpn. **76** (2007) 023708.
3) Y. Yasui, Y. Naito, K. Sato, T. Moyoshi, M. Sato, and K. Kakurai: J. Phys. Soc. Jpn. **77** (2008) 023712.
4) S. Park, Y. J. Choi, C. L. Zhang, and S-W. Cheong: Phys. Rev. Lett. **98** (2007) 057601.
5) S. Seki, Y. Yamasaki, M. Soda, M. Matsuura, K. Hirota, and Y. Tokura: Phys. Rev. Lett. **100** (2008) 127201.
6) Y. Kobayashi, K. Sato, Y. Yasui, T. Moyoshi, M. Sato, and K. Kakurai: J. Phys. Soc. Jpn. **78** (2009) 084721.
7) Y. Yasui, K Sato, Y. Kobayashi, and M. Sato: J. Phys. Soc. Jpn. **78** (2009) 084720.
8) H. Katsura, N. Nagaosa, and A. V. Balatsky: Phys. Rev. Lett. **95** (2005) 057205.
9) M. Mostovoy: Phys. Lev. Lett. **96** (2006) 067601.
10) M. Sato, Y. Yasui, Y. Kobayashi, K. Sato, Y. Naito, Y. Tarui, and Y. Kawamura: Solid State Sciences, available online 10 March 2009.
11) T. Kimura, T. Goto, H. Shintani, K. Ishizaka, T. Arima, and Y. Tokura: Nature **426** (2003) 55.
12) T. Goto, T. Kimura, G. Lawes, A. P. Ramirez, and Y. Tokura: Phys. Rev. Lett. **92** (2004) 257201.
13) T. Kimura, G. Lawes, and A. P. Ramirez: Phys. Rev. Lett. **94** (2005) 137201.
14) B. Kundys, A. Maignan, and Ch. Simon: Appl. Phys. Lett. **94** (2009) 072506.
15) A. W. Mombrú, K. Prassides, C. Christides, R. Erwin, M. Pissas, C. Mitros, and D. Niarchos: J. Phys.: Condens. Matter **10** (1998) 1247.
16) Y. Yasui *et al.*: unpublished.
17) A. W. Mombru, A. E. Goeta, H. Pardo, P. N. Lisboa-Filho, L. Suescun, R. A. Mariezcurrena, O. N. Ventura, K. H. Andersen, and F. M. Araujo-Moreira: J. Solis State Chem. **166** (2002) 251.
18) O. Heyer, N. Hollmann, I. Klassen, S. Jodlauk, L. Bohatý, P. Becker, J. A. Mydosh, T. Lorenz, and D. Khomskii: J. Phys.: Condens. Matter **18** (2006) L471.
19) J. Bernhard, B. Lebech, and O. Beckman: J. Phys. F: Met. Phys. **14** (1984) 2379.
20) S. Iikubo, Y. Ohno, Y. Kobayashi, Y. Yasui, M. Ito, M. Soda, M. Sato, and K. Kakurai: J. Phys. Soc. Jpn. **73** (2004) 1023.
21) V. Caignaert, I. Mirebeau, F. Bouree, N. Nguyen, A. Ducouret, J-M. Greneche, and B. Raveau: J. Solid State Chem. **114** (1995) 24.

## Figure Caption

Fig. 1. As an example of the X-ray powder patterns of $RBaCuFeO_5$, the data obtained for $TmBaCuFeO_5$ with CuKα radiation are shown. In the right part, the structure of $RBaCuFeO_5$ is schematically shown.

Fig. 2. (a) Magnetic susceptibility χ of $YBaCuFeO_5$ measured with *H*=1 T is shown against *T*. (b) The dielectric susceptibility ε of $YBaCuFeO_5$ is shown against *T*. $T_{N1}$ and $T_{N2}$ are the temperatures of the susceptibility peaks at *H*=1 T. In (c) and (d), the pyrocurrent $I_p$ and electric polarization *P* are shown, respectively, against *T*. To the data curves, the voltages between the electrodes of the plate-like sample applied during the cooling processes are attached (The measurements were carried out at zero applied voltage.) In the figures, $T_{N1}$ and $T_{N2}$ are the temperatures of the susceptibility peaks at *H*=0, and $T_{N2}$' is the peak temperature of d$\chi$/d*T* at $H \to 0$.

Fig. 3. (a) *T* dependences of the electric polarization *P* of $LuBaCuFeO_5$ obtained by measuring $I_p$ with increasing *T* at zero applied electric field. The voltages between the electrodes of the plate-like sample applied during the cooling processes are indicated. The temperatures $T_{N2}$, $T_{N2}$' and $T_0$ at $H \to 0$ are indicated by the arrows. In (b), d$\chi$/d*T* data obtained at *H*=1, 3 and 5.5 T (***H***||*ab*) for a *c*-axis aligned sample of $LuBaCuFeO_5$ (in stycast) are shown against *T*. The temperatures $T_{N2}$' are indicated by the arrows for various ***H*** applied within the *ab* plane. Inset shows the *T* dependence of the magnetic susceptibility obtained at *H*=1 T (***H***||*ab*). $T_{N2}$ at *H*=1 T is indicated by the arrow.

Fig. 4. (a) *T* dependence of the electric polarization obtained for $TmBaCuFeO_5$ is shown, where the voltage between the electrodes of the plate-like sample applied during the cooling processes are attached.

Fig. 1

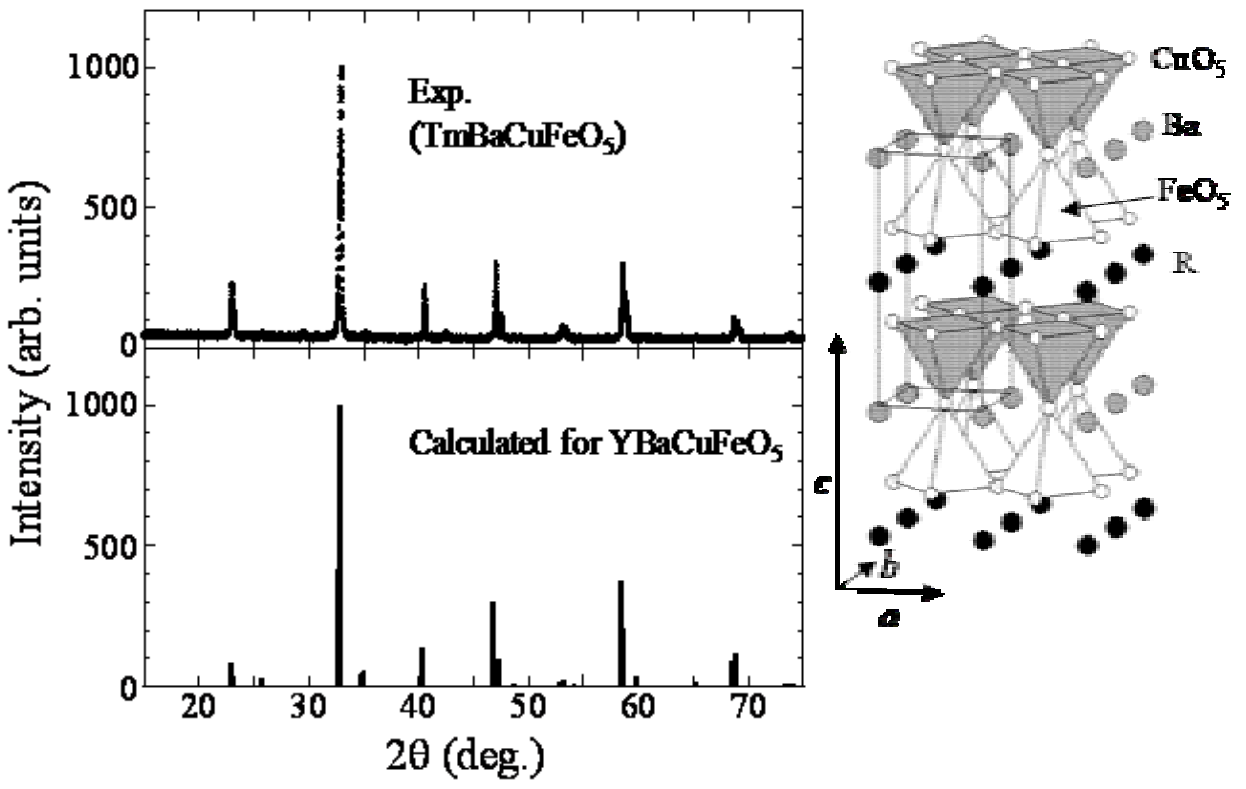

Fig. 2

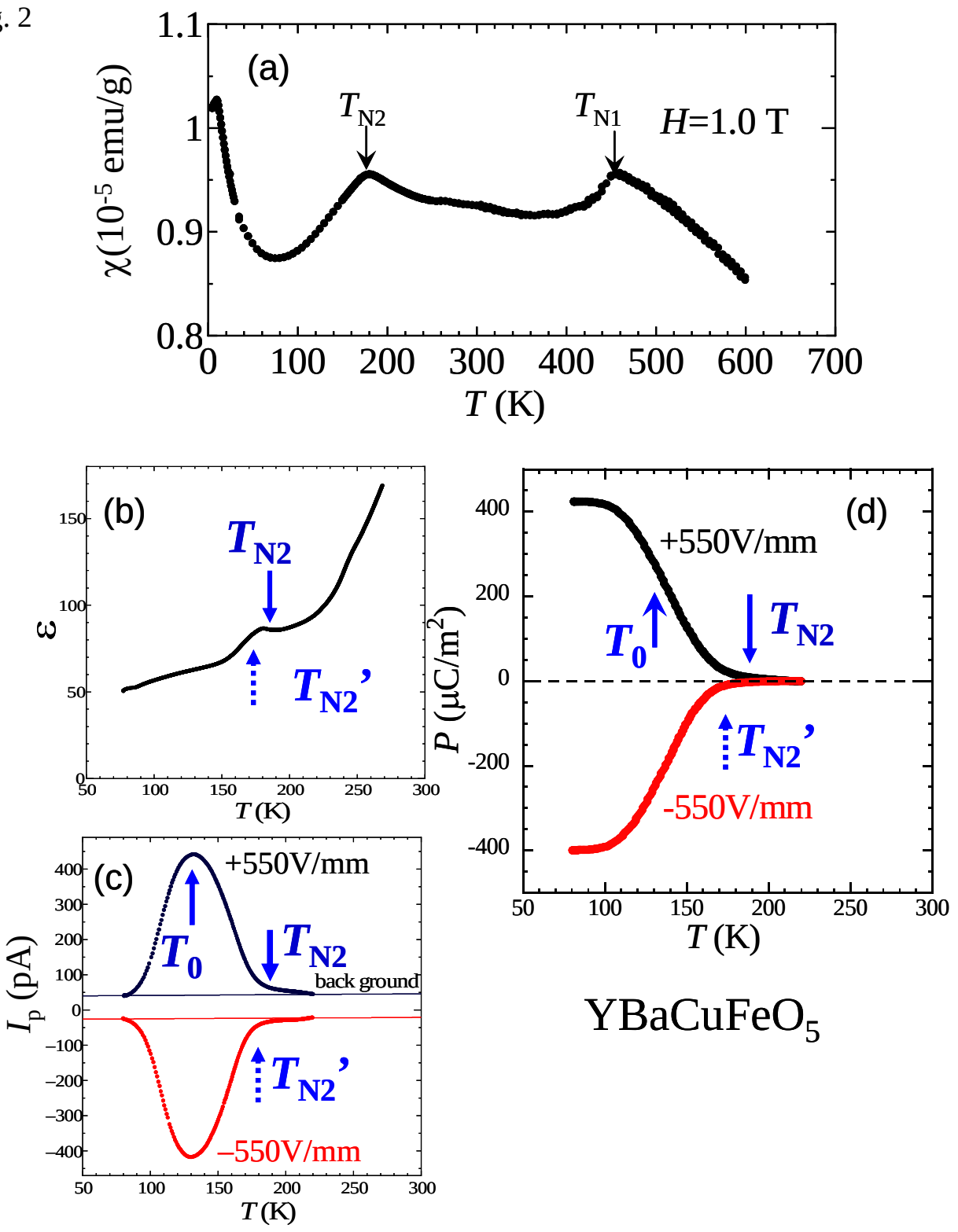

Fig.4

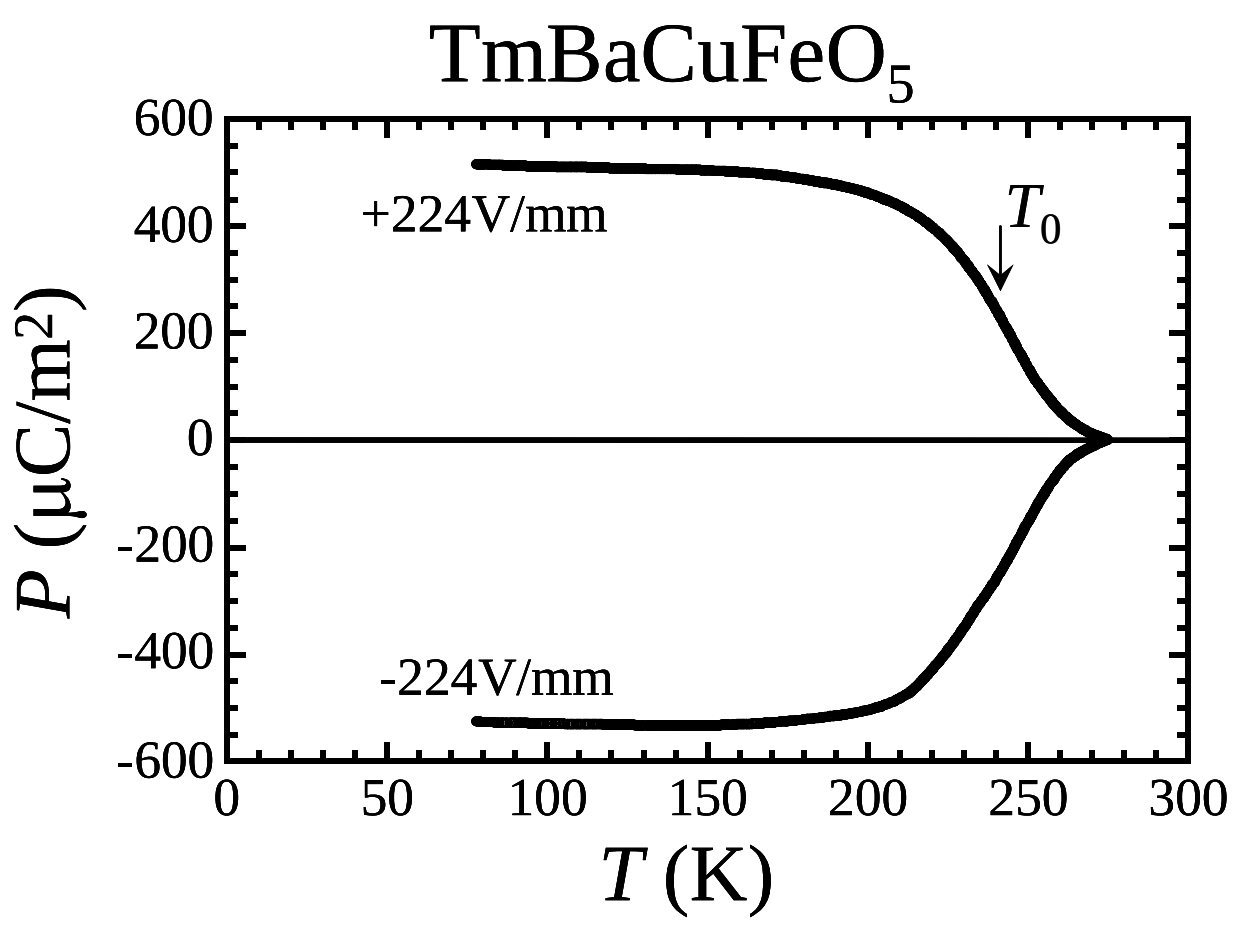

Fig. 3

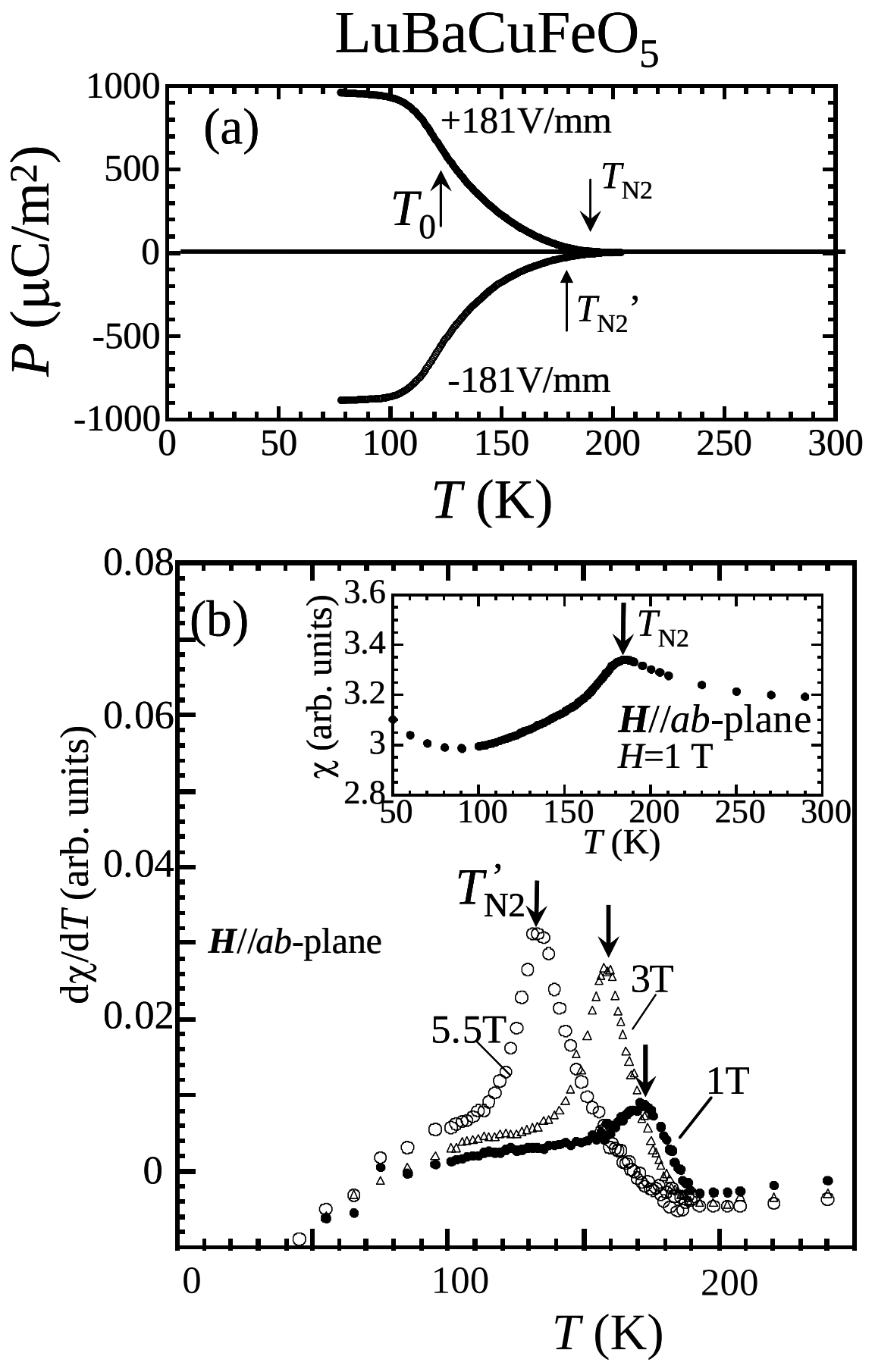